\documentclass[twocolumn, prd, amssymb]{revtex4}
\def\CH{{\cal H}}

\def\IZ{\relax\ifmmode\mathchoice
{\hbox{\cmss Z\kern-.4em Z}}{\hbox{\cmss Z\kern-.4em Z}}
{\lower.9pt\hbox{\cmsss Z\kern-.4em Z}}
{\lower1.2pt\hbox{\cmsss Z\kern-.4em Z}}\else{\cmss Z\kern-.4em }\fi}
\def\IC{\relax\hbox{$\inbar\kern-.3em{\rm C}$}}
\def\IR{\relax{\rm I\kern-.18em R}}
\def\bZ{{\bf Z}}

\def\bC{{\bf C}}

\def\bR{{\bf R}}

\def\r{\rho}

\def\d{\delta}

\def\p{\pi}

\def\1{\relax 1 { \rm \kern-.35em I}}

\def\frac#1#2{{#1 \over #2}}

\def\p+{{\partial_+}}

\def\apm{\alpha^{\prime}}

\def\[{\left [}
\def\]{\right ]}
\def\({\left (}
\def\){\right )}

\def\ajou#1&#2(#3){\ \sl#1\bf#2\rm(19#3)}
\def\npb#1#2#3{{\sl Nucl. Phys.} {\bf B#1} (#2) #3}
\def\plb#1#2#3{{\sl Phys. Lett.} {\bf B#1} (#2) #3}
\def\prl#1#2#3{{\sl Phys. Rev. Lett. }{\bf #1} (#2) #3}
\def\prd#1#2#3{{\sl Phys. Rev. }{\bf D#1} (#2) #3}

\def\cmp#1#2#3{{\sl Comm. Math. Phys. }{\bf #1} (#2) #3}
\def\mpl#1#2#3{{\sl Mod. Phys. Lett. }{\bf #1} (#2) #3}

\def\TrH#1{ {\raise -.5em
                      \hbox{$\buildrel {\textstyle  {\rm Tr } }\over
{\scriptscriptstyle \CH _ {#1}}$}~}}

\begin{document}
\preprint{HUPT-01/A051}
\preprint{TIFR/TH/01-42}
\preprint{hep-th/0111004}
\title{Tachyon Condensation and Black Hole Entropy}
\author{Atish Dabholkar$^{a,b}$}
\affiliation{$^a$ Jefferson Physical Laboratory,
  Harvard University,
  Cambridge, MA 02138, USA \\
$^b$ Department of Theoretical Physics,  
Tata Institute of Fundamental Research, Mumbai 400005, India}

\date{October 2001}

\begin{abstract}
String propagation on a cone with deficit angle $2\pi(1-{1\over N})$
is considered for the purpose of computing the entropy of a large mass
black hole. The entropy computed using the recent results on
condensation of twisted-sector tachyons in this theory is found to be
in precise agreement with the Bekenstein-Hawking entropy.
\end{abstract}

\maketitle

\centerline{\it Introduction}
\medskip
The statistical interpretation of Bekenstein-Hawking entropy of a
black hole remains an outstanding problem in quantum gravity. For a
black hole in $d$ spacetime dimensions,  the entropy
$S$ is given by a universal formula
\begin{equation}%
S = {A \over 4 G \,}
\label{BH}
\end{equation}
that depends only on the area $A$ of the event horizon and the
$d$-dimensional Newton's constant $G$. Thermodynamically, this entropy
behaves in every respect like ordinary entropy and unites the second
law of thermodynamics with the area theorems of classical general
relativity into an elegant generalized second law
\cite{Beke,Hawk}. These beautiful results demand that, like any
other entropy, the black hole entropy also must have a statistical
interpretation in terms of underlying microstates.

Recent progress in string theory has shown that this is indeed true
for a large class of supersymmetric black holes. The microstates of
these special black holes are precisely countable and can completely
account for their entropy \cite{Sen,StVa}.  These striking results can be
extended to near-extremal black holes as well as to certain
nonsupersymmetric charged black holes. At present, however, these
methods cannot be applied to the more general case of a
nonsupersymmetric neutral black hole. 

We will address the problem of microstates of a Schwarzschild black
hole from a very different perspective based on earlier ideas of 't
Hooft \cite{Hoof}, Susskind \cite{Suss}, and others \cite{ZuTh, BKLS,
Sred, CaWi}.  't Hooft \cite{Hoof} has advocated that the thermal
entropy of the heat bath seen by a Schwarzschild observer should
account for the Bekenstein-Hawking entropy.  This would then offer the
desired statistical interpretation of the entropy in terms of the
near-horizon microstates of the heat bath.  In field theory, the
leading contribution to the thermal entropy is quadratically divergent
in the ultraviolet \cite{Hoof} and is proportional to the area of the
horizon.  If the cut-off is of the order of the Planck length, then
the thermal entropy is of the right order of magnitude to be
identified with the Bekenstein-Hawking entropy.

In the context of field theory, there are several difficulties with
this appealing idea.  For example, the thermal entropy depends on an
arbitrary cutoff and its precise identification with the Planck length
is not clear. The thermal entropy depends on the species and the
couplings of the various particles in the theory whereas the black
hole entropy is species-independent. Finally, since the thermal
entropy always starts at one loop, it
is difficult to see how it can possibly account for the tree level
black hole entropy which is inversely proportional to the coupling
constant.  't Hooft has argued that it is necessary to understand the
ultraviolet structure of the theory to address these questions and
therefore these difficulties will be resolved in the correct
short-distance theory of quantum gravity.

In string theory, ultra-violet divergences are expected to be
appropriately controlled and Susskind, in particular, has argued that
string theory offers a suitable framework for realizing this proposal
\cite{Suss}. We will pursue these ideas further by
considering string theory on the near horizon geometry in Euclidean
formalism following earlier work in \cite{Dabh,DabhI}.

\medskip
\centerline{\it Strings on a Cone}
\medskip
Consider a Schwarzschild black hole in four dimensions with a large mass $M$.
The metric is given by 
\begin{equation}%
ds^2 = -(1-{2GM\over r})dt^2 + (1-{2GM\over r})^{-1} dr^2 + r^2 d\Omega^2
\label{metric}
\end{equation}
We are interested in the thermal entropy of string modes in this
background near the horizon as seen by a Schwarzschild observer. 
To focus on the near-horizon region we choose
coordinates $\r = \sqrt{8GM(r-2GM)}$ and $\eta={l t
\over 2GM }$ in which the metric becomes
\begin{equation}%
ds^2 = -({\r\over l})^2 d\eta^2 + d\r^2 + (2GM)^2 d\Omega^2.
\label{near}
\end{equation}
Here $l$ is an arbitrary parameter that sets the length scale which in
our context will be taken to be the string length. We see from the
form of the metric that $\r$ measures the proper distance from the
horizon at $\rho =0$ and $\eta$ is the proper time measured by a
Schwarzschild observer located at proper distance $\rho =l$. We will
refer to this location as the `stretched horizon'. Energy at the stretched
horizon is  related to asymptotic energy by the red-shift factor $l/4GM$.

In the limit $GM >> l$, the two-sphere at the horizon can be
approximated by a flat 2-dimensional transverse space and then the
space has flat Minkowski geometry. Thus, in this limit, the
Schwarzschild observers are exactly like Rindler observers in uniform
acceleration in Minkowski space. More generally, for a d-dimensional
black hole the near horizon geometry is given by the $d$-dimensional
Rindler metric with topology ${\bR^2}\times {\bR^{d-2}}$
\begin{equation}%
ds^2 = -\r^2 d\eta^2 + d\r^2 + \sum_{i=1}^{d-2} (dx^i)^2,
\label{rindler}
\end{equation}
where $x^i$ are the coordinates of the $(d-2)$-dimensional transverse
space and we have chosen units to set $l=1$.

The observer at the stretched horizon sees a thermal bath at Rindler
temperature $T={1\over 2\pi }$ which, as usual, is inversely
proportional to the periodicity of Euclidean time. Under Euclidean
continuation, $\eta = -i \theta$, the metric in the $\bR^2$ factor
is given by the flat metric in polar coordinates
\begin{equation}%
ds^2 = \r^2 d\tau^2 + d\r^2
\label{polar}
\end{equation}%
The Euclidean time $\tau$ is thus an angular variable and when its
periodicity is $2\pi$ corresponding to the Rindler temperature, the
metric is smooth at the origin. The Rindler temperature at the
stretched horizon is red-shifted to the Hawking temperature
$T_H={1\over 8\pi G M}$ seen by the asymptotic observer.

The partition function $Z$ of this thermal bath in string theory would
be given by a functional integral of all string modes on the Euclidean
Rindler space.  The thermal entropy of this heat bath $S$ would be
given, as usual, by $S= -{\partial F \over \partial T}$ as a
derivative of the free energy, $F \equiv - T\ln Z$. Therefore, to
calculate the entropy, we need to find the variation in the free
energy of a string gas to order $\delta$ as we vary the temperature seen
by the observer at the stretched horizon to ${1\over 2\pi}
+\d$. Changing the temperature changes the periodicity of Euclidean
time in (\ref{polar}) to $ 2\pi -\delta$ to leading order. The Euclidean
geometry is now conical with deficit angle $\delta$ and with a
curvature singularity at the tip of the cone.

One of the difficulties in evaluating the partition function for
strings on a cone with arbitrary deficit angle is that the cone is not
a solution of the string equations of motion. The Einstein equation is
not satisfied unless there is an explicit source at the tip to account
for the curvature.  One would thus require an off-shell formulation of
string theory to evaluate this partition function.  

We will instead proceed differently. We will consider string theory on
a cone obtained as an orbifold $\bC /\bZ_N$ in
\cite{Dabh,DabhI,LoSt}. This corresponds to the temperature $T =
{N\over 2\pi}$ at the stretched horizon and the deficit angle is
$\delta =2\pi (1-{1\over N})$. Because the orbifold is a conformal
field theory, for these special values of the deficit angle labeled by
an integer, the tree-level string equations of motion are indeed
satisfied. In addition to the bulk modes, there are states from $N-1$
twisted sectors labeled by $k=1, \ldots, N-1$. We will take both $k$
and $N$ to be odd to simplify the subsequent discussion. The ground
state in each sector is tachyonic and its mass is given by $\apm m^2 =
-2(1-k/N)$. In the next section we evaluate the free energy $F(N)$ as
a function of $N$ by properly taking into account infrared divergences
due to the tachyons and then compute the entropy analytically
continuing in $N$.

\medskip
\centerline {\it Tachyon Condensation and Black Hole Entropy}
\medskip

The one-loop partition function for strings on the orbifold
$\bC/\bZ_N$ can be easily written down. It is modular invariant and
therefore ultraviolet finite as expected for a string partition
function. This is already an improvement over the field theory
calculation which was UV divergent.  However the partition 
function has severe infrared divergences because of the tachyons.

Infrared divergences, unlike ultraviolet divergences, are not a matter
of renormalization but signify important physics. In this context, the
existence of tachyons in the spectrum suggests that the thermal
ensemble is unstable. It was speculated in \cite{Dabh,DabhI} that
these tachyons can condense causing a phase transition and  
the latent heat of this transition could account for the tree level
black hole entropy. However, in the absence of a good candidate for
the endpoint of tachyon condensation, it was not possible to pursue
this further.

Recent work of Adams, Polchinski, and Silverstein \cite{APS} has
provided new insights about condensation of these tachyons.  They have
argued that tachyon condensation relaxes the cone to flat space. The
most convincing evidence for this claim comes from the geometry seen
by a D-brane probe in the sub-stringy regime. In the probe theory, one
can identify operators with the right quantum numbers under the quantum
$\bZ_N$ symmetry of the orbifold that correspond to turning on
tachyonic vevs. By selectively turning on specific tachyons, the
quiver theory of the probe can be `annealed' to successively go from
the $\bZ_N$ orbifold to lower $\bZ_{N-2}$ orbifold all the way to flat
space. The deficit angle seen by the probe in this case changes
appropriately from $2\pi(1- {1\over N})$ to $2\pi (1- {1\over N-2})$.

These results are consistent with the assumption that in the field
space of tachyons there is a potential $V(T)$ where we collectively
denote all tachyons by $T$. The $\bZ_N$ orbifold sits at the top of
this potential, the various $\bZ_M$ orbifolds with $ M < N$ are the
other extrema of this tachyonic potential, and flat space is at the
bottom of this potential.  Such a potential can also explain why a
conformal field theory exists only for special values of deficit
angles. We will be concerned here with the static properties such as
the endpoint of tachyon condensation and the effective height of the
tachyon potential and not so much with dynamical details of the
process of condensation.

Let us return now to the computation of the black hole entropy. We
would like to evaluate the free energy of a $\bZ_N$ orbifold as a
function of $N$. Now, the existence of tachyons in the orbifold
implies that we are expanding the string field theory functional
integral around a maximum. The asymptotic expansion provided by string
perturbation theory around this point is as a result IR divergent and
essentially useless. To correctly evaluate the partition function, we
must expand around the stable saddle point at the minimum of the
potential. The leading semiclassical contribution to the partition
function will be given by $Z \sim \exp{-S_E}$ where $S_E$ is the
classical Euclidean action after condensation. To elaborate this point
let us consider a toy model in field theory of a single scalar field
$\phi$ with double well potential $-m^2 \phi^2 + g^2
\phi^4$. The perturbative expansion for the
partition function around $\phi =0$ is IR divergent which signifies
that we have expanded around the wrong saddle point. The stable saddle
point is at $\phi^2 = {m^2 \over 2g^2}$ and the leading semiclassical 
contribution to the partition function will be given by $Z \sim
\exp{-S_E}$ where $S_E = {-m^4 \over 4g^2}$ 
is the change in the classical Euclidean action after condensation.

It may seem difficult to evaluate the change in the classical action
between the $\bZ_N$ orbifold and flat space but we are helped by the
fact that, for the orbifold conformal field theory, the equations of
motion for the dilaton and the graviton are satisfied exactly.  To
extract this information, let us consider the Lorentzian string
effective action, for concreteness, first to leading order in $\apm$
\begin{eqnarray}%
S = {1\over 16\pi G} \int_M \sqrt{-g}\, e^{-2\phi}\, [R +4 (\nabla \phi)^2
-\delta^2(x)V(T)]{\nonumber}\\
+{1\over 8\pi G} \int_{\partial{M}} \sqrt{-g}\,e^{-2\phi}\, K\, ,\qquad
\label{action}
\end{eqnarray}%
where $K$ is the extrinsic curvature and $\delta^2 (x) V(T)$ denotes
the tachyon potential localized at the tip of the cone. The extrinsic
curvature term is as usual necessary to ensure that the effective
action reproduces the string equations of motion for variations
$\delta\phi$ and $\delta g$ that vanish at the boundary.

The action is very similar to the one for a cosmic string in four
dimensions.  The tachyon potential supplies an 8-brane source term for
gravity. Einstein equations imply $R=\delta^2(x)V(T)$ and therefore a
conical curvature singularity at $x=0$. Because of this equality,
there is no source term for the dilaton and as a result the dilaton
equations are satisfied with a constant dilaton. We see that the bulk
contribution to the action is zero for the solution. The boundary has
topology $\bR^8\times {\bf S}^1$. For a cone, the circle ${\bf S}^1$ has
radius $r$ but the angular variable will go from $0$ to $2\pi \over
N$. The extrinsic curvature for the circle equals $1/r$ and thus the
contribution to the action from the boundary term equals $A \over 4 G
N$. We have to remember a factor of $-i$ in Euclidean continuation of
$\sqrt{-g}$.

In the conformal field theory, we should worry about the higher order $\apm$
corrections to the effective action. These corrections are dependent on field
redefinitions or equivalently on the renormalization scheme of the
world-sheet sigma model. However, the total contribution of these
corrections to bulk action must nevertheless vanish for the orbifold
because we know that the equations of motion of the dilaton are
satisfied with a constant dilaton which implies no source terms for
the dilaton in the bulk. Thus, the entire contribution to the action
comes from the boundary term even when the $\apm$ corrections are
taken into account and we can reliably calculate it in a scheme
independent way using the conical geometry of the exact solution at
the boundary.

{}From the change in the classical action between the $\bZ_N$
orbifold and flat space, we can calculate the leading semi-classical
contribution to the free energy up to an $N$-independent additive
constant:
\begin{equation}%
F(N) = -{A\over 4G} {(N-1)\over 2\pi}
\label{free}
\end{equation}%
The resulting entropy is 
\begin{equation}%
S = -2\pi {\partial{F(N)} \over \partial{N}} = {A \over 4G}\, , 
\label{final}
\end{equation}%
which is in precise agreement with the Bekenstein-Hawking entropy.
Note that the entropy is independent of $N$ even though the latent heat
scales with $N$.

Unlike the change in the classical action, the height of the tachyon
potential is dependent on the renormalization scheme. It seems likely,
however, that there exists a particular renormalization scheme of the
$N=2$ supersymmetric sigma model of the cone in which the beta
function equations are reproduced exactly by (\ref{action}) to all
orders in $\apm$ and not just to leading order. This would be
analogous to Calabi-Yau compactifications where one can argue that
there always exists a scheme in which the metric is Ricci flat
\cite{NeSe}. In such a renormalization scheme, the dilaton would be constant,
the metric would be precisely conical as for the exact solution of the
orbifold. In this scheme, the height of the tachyon potential $\mu_N$
for the $\bZ_N$ would then be directly related to the deficit angle
$\delta$ of the solution using Einstein equations from (\ref{action})
would be given by
\begin{equation}%
\mu_N= 2 {\delta} = 4\pi (1- {1\over N}).
\label{height}
\end{equation}%
It would be interesting to verify this prediction in some form of
closed string field theory if the scheme-dependence can somehow be
taken into account.

\bigskip
\centerline{\it Conclusions and Discussion}
\medskip
We end with a few comments and open problems.

The calculation here may seem similar to the calculation of Gibbons
and Hawking or Teitelboim \cite{GiHa,Teit} where the entropy comes
from the classical gravitational action. However, conceptually, it is
fundamentally different.  The Gibbons and Hawking calculation in
canonical gravity is in an effective low energy theory and is
insensitive to the high energy modes. In canonical gravity, a cone is
not a saddle point.  One can introduce energy density $\mu$ as a
chemical potential for holding the area of the horizon fixed and then the cone
of arbitrary deficit angle can be a saddle point \cite{Teit,Suss}. But
this chemical potential is not related to the potential of any
dynamical field. There are no IR divergences in these calculations and
the UV divergences simply renormalize Newton's constant in the
effective theory.

Somehow, string theory, as a more complete theory of gravity,
automatically supplies additional tachyonic fields at the Euclidean
horizon. The string partition function is a functional integral over
the bulk fields such as the metric as well as the twisted fields
including the tachyons and thus is clearly different from the Gibbons
and Hawking path integral. The cone is now a saddle point of the full
string equations of motion only for integer $N$ but without having to
introduce any chemical potential by hand.  The UV finite partition
function is now IR divergent, which implies a dynamical
instability. One of the main motivations of this paper is to take
seriously the dynamics of the new fields supplied by string
theory. This dynamics at the tip seems to capture the physics of the
heat bath close to the horizon. The entropy computed this way is
universal in that it does not depend on what bulk theory we started
with but is determined entirely by the potential of the tachyonic
degrees of freedom near the horizon.

The UV divergence of entropy in field theory is intimately related to
the puzzle of loss of information in black hole evaporation.  If the
entropy does have a statistical interpretation in terms of counting of
states, then its divergence would suggest an infinite number of states
associated with a finite mass black hole.  As long as the black hole
has an event horizon, it can apparently store an arbitrary amount of
information in terms of correlations between the outgoing radiation
and the high energy modes near the horizon. When the horizon
eventually disappears, the information in these correlations is
irretrievably lost. Finiteness of the entropy as we have found both in
UV and IR, on the other hand, implies unitary evolution without
information loss.

The tachyons at the tip of the cone are very similar to the thermal
tachyons in string theory at finite temperature that come from the
winding modes around the Euclidean time direction
\cite{Sath,Koga}. They cannot be interpreted as states in the spectrum
but are rather new order-parameter fields. In the Lorentzian
continuation, the tachyon condensate near the tip of the cone would
seem to imply a Hagedorn-like phase near the Lorentzian horizon
reminiscent of the `membrane paradigm' \cite{TPM}. Apparently, the
entropy of this phase accounts for the black hole entropy. A useful
analogy is the deconfinement transition in large $N$ QCD. It is as
though we are able to identify the order parameter for the
deconfinement transition in the form of tachyons and compute the
latent heat of the transition precisely from the potential to find a
tree level contribution of order $N^2$ that is suggestive of gluon
degrees of freedom. It would be desirable to figure out in a
Hamiltonian formalism, the analog of the gluon degrees of
freedom. Since the thermal entropy accounts for the entire tree-level
entropy, it suggests that the Einstein action is wholly induced by the
interactions of the UV degrees of freedom. Matrix theory \cite{BFSS}
or a gauge theory dual \cite{AGMO} may be the right framework to
address this question. It is interesting to explore the tachyon
landscape irrespective of the problem of black hole entropy and to
verify the form of the potential in a closed string field theory. It
would also be interesting to pursue the implications of these results
for the Hagedorn transition in string theory \cite{AtWi} as well as
for holography \cite{Hoof1, Suss1, AGMO}.

\medskip
\centerline{\it Acknowledgments}
I am grateful to Rajesh Gopakumar, Kentaro Hori, and Sandip Trivedi
for many valuable discussions.
\medskip

\begingroup\raggedright

\endgroup
\end{document}